\documentclass[12pt]{article}
\usepackage{a4wide}
\usepackage{latexsym}

\def\bq{\begin{eqnarray}}
\def\eq{\end{eqnarray}}
\def\l{\langle}
\def\r{\rangle} 
\def\eps{\varepsilon}

\newlength{\dinwidth} \newlength{\dinmargin}
\begin{document}

\thispagestyle{empty}

\begin{flushright}
  NIKHEF-2001-002\\
  UPRF-2001-04
\end{flushright}

\vspace{1.5cm}

\begin{center}
  {\Large\bf Dipole formalism with heavy fermions\\}
  \vspace{1cm}
  {\sc L. Phaf$^1$ and  S. Weinzierl$^2$}\\
  \vspace{1cm}
  {\it $^1$ NIKHEF Theory Group\\
       Kruislaan 409, 1098 SJ Amsterdam, The Netherlands,} \\
\vspace*{4mm}
  {\it $^2$ Dipartimento di Fisica, Universit\`a di Parma,\\
       INFN Gruppo Collegato di Parma, 43100 Parma, Italy} \\
\end{center}

\vspace{2cm}

\begin{abstract}\noindent
  {We extend the dipole formalism of Catani and Seymour to QCD
   processes involving heavy fermions.
   We give the appropriate subtraction terms together with their
   integrated counterpart. All calculations are done within dimensional
   regularization. Our formulae can be used with three variants of
   dimensional regularization (conventional dimensional regularization,
   the 't Hooft-Veltman scheme or the four-dimensional scheme).
   }
\end{abstract}

\vspace*{\fill}

\newpage

\reversemarginpar

\section{Introduction}

Studies of hard-scattering processes in QCD with heavy particles
are important at present or future colliders.
Reliable theoretical prediction require at least the evaluation 
of next-to-leading order (NLO) QCD corrections.
Next-to-leading order calculations have to combine virtual one-loop
corrections with the real emission contributions from unresolved partons.
Treated separately, each of the two parts gives an infrared divergent
contribution. Only the sum of the virtual corrections and the real
emission contributions is infrared finite.
Setting up a numerical general-purpose NLO Monte-Carlo program requires
therefore the analytical cancellation  
of infrared singularities before any numerical integration can be done.
The two main methods to handle this task are phase space slicing 
\cite{Fabricius:1981sx}-\cite{Kramer:1989mc}
and the subtraction method
\cite{Ellis:1981wv,Kunszt:1989km}
.
For massless partons both methods are available and have been applied to a variety of specific processes, see for example ref. \cite{Baer:1989jg}-\cite{Kunszt:1992tn} and the references therein.
For electron-positron annihilation a general formulation of phase space slicing
has been given by Giele and Glover \cite{Giele:1992vf}. It has been
extended to initial-state partons by these authors and Kosower
\cite{Giele:1993dj}.
The extension of phase space slicing to massive partons and identified hadrons
has been given by Laenen and Keller \cite{Keller:1998tf}.\\

There are two general formulations of the subtraction method. One is the residue approach 
by Frixione, Kunzst and Signer \cite{Frixione:1996ms}
and the other the dipole formalism by Catani and Seymour \cite{Catani:1997vz},
both variants can handle massless partons in the final and/or initial state.
The subtraction method has already been applied to some specific processes with massive partons 
\cite{Mangano:1992jk}-\cite{Harris:1996vd}.
Up to now there is no extension of the dipole formalism to massive partons.
Dittmaier has considered the dipole formalism for photon radiation 
from fermions \cite{Dittmaier:1999mb}. In that work infrared divergences are
regularized with small masses (as it is popular in electroweak physics).
However this does not allow a simple application to QCD, where divergences
are usually regularized by dimensional regularization.\\

In this paper we extend the dipole formalism
to heavy fermions.
The formulae we provide are relevant to top, bottom and charm production.
With a simple change of the colour factors they can be used as well for gluino production in
SUSY QCD.
Our results are not applicable for
processes with identified hadrons in the final state,
massive initial-state partons and processes with different species of massive fermions of unequal
masses. \\

One-loop amplitudes may be calculated in different
variants of dimensional regularization, such as conventional dimensional regularization
(CDR), the 't Hooft - Veltman scheme (HV) or the four-dimensional scheme (FD), whereas
the Born amplitudes entering the real emission part and/or parton densities
are given in another variant.
Of course, the final result has to be scheme-independent. The obvious way to ensure this is to
calculate everything in the same scheme. 
Often this is not the most
economical solution. Loop amplitudes are most easily computed in the FD scheme, whereas
parton densities are given in the conventional scheme.
Within the dipole formalism it is possible to perform different parts of the 
calculation with different variants of dimensional regularization and to 
correct for the mismatch by some universal terms \cite{Catani:1997pk}.
We would like to keep this freedom also in the extension to massive partons.
We therefore keep track of all scheme-dependent terms
and our final results can be used with any variant of
dimensional regularization (CDR, HV or FD). \\

Recently, Catani, Dittmaier and Trocsanyi considered the singular behaviour
of one-loop QCD and supersymmetric QCD amplitudes with massive partons in the dipole framework \cite{Catani:2000ef}.
As a byproduct, we confirm their results for the singular terms.\\

This paper is organized as follows. In the next section we review the
dipole formalism and the factorization of QCD amplitudes in the soft
and collinear limit.
In section 3 we outline our calculational technique.
In section 4 we derive the $D$-dimensional dipole phase space measure.
Section 5 gives the dipole terms together with the integrated counterpart
if all particles are in the final state.
In section 6 we consider the case if there are QCD partons in the initial
state.
Finally, section 7 contains our conclusions.

\section{The dipole formalism and factorization in singular limits}

In this section we briefly review the dipole formalism and the factorization properties
of QCD amplitudes in the soft and collinear limit.
We use the notation of Catani and Seymour \cite{Catani:1997vz}.\\
\\
The dipole formalism is based on the subtraction method. 
We explain it for electron-positron annihilation, where all QCD partons are in the final state.
The NLO cross section is rewritten as
\begin{eqnarray}
\sigma^{NLO} & = & \int\limits_{n+1} d\sigma^R + \int\limits_n d\sigma^V \nonumber \\
& = & \int\limits_{n+1} \left( d\sigma^R - d\sigma^A \right) + \int\limits_n \left( d\sigma^V + \int\limits_1 
d\sigma^A \right).
\end{eqnarray}
In the second line an approximation term $d\sigma^A$ has been added and subtracted.
The approximation $d\sigma^A$ has to fulfill the following requirements:
\begin{itemize}
\item $d\sigma^A$ must be a proper approximation of $d\sigma^R$ such as to have the same pointwise singular behaviour
(in $D$ dimensions) as $d\sigma^R$ itself. Thus, $d\sigma^A$ acts as a local counterterm for $d\sigma^R$ and one
can safely perform the limit $\varepsilon \rightarrow 0$. This defines a cross-section contribution
\begin{eqnarray}
\sigma^{NLO}_{\{n+1\}} & = & \int\limits_{n+1} \left( \left. d\sigma^R \right|_{\varepsilon=0}
- \left. d\sigma^A \right|_{\varepsilon=0} \right).
\end{eqnarray}
\item Analytic integrability (in $D$ dimensions) over the one-parton subspace leading to soft and collinear
divergences. This gives the contribution
\begin{eqnarray}
\sigma^{NLO}_{\{n\}} & = & \int\limits_n \left( d\sigma^V + \int\limits_1 d\sigma^A \right)_{\varepsilon=0}.
\end{eqnarray}
\end{itemize}
The final structure of an NLO calculation is then
\begin{eqnarray}
\label{NLOcrosssection}
\sigma^{NLO} & = & \sigma^{NLO}_{\{n+1\}} + \sigma^{NLO}_{\{n\}}.
\end{eqnarray}
Since both contributions on the r.h.s of eq.(\ref{NLOcrosssection}) are now finite,
they can be evaluated with numerical methods.
The $(n+1)$ matrix element is approximated by a sum of dipole terms
\begin{eqnarray}
\lefteqn{\sum\limits_{pairs\; i,j} \sum\limits_{k \neq i,j} {\cal D}_{ij,k} =} \nonumber \\
& = & 
\sum\limits_{pairs\; i,j} \sum\limits_{k \neq i,j} 
- \frac{1}{2 p_i \cdot p_j}
\langle 1, ..., \tilde{(ij)},...,\tilde{k},...|
\frac{{\bf T}_k \cdot {\bf T}_{ij}}{{\bf T}^2_{ij}} V_{ij,k} |
1,...,\tilde{(ij)},...,\tilde{k},... \rangle, \nonumber \\
\end{eqnarray}
where the emitter parton is denoted by $\tilde{ij}$ and the spectator
by $\tilde{k}$.
Here ${\bf T}_i$ denotes the colour charge operator \cite{Catani:1997vz} for parton $i$ and
$V_{ij,k}$ is a matrix in the helicity space of the emitter with the correct
soft and collinear behaviour.
$|1,...,\tilde{(ij)},...,\tilde{k},... \rangle$ is a vector in colour- and helicity space.
By subtracting from the real emission part the fake contribution we obtain
\begin{eqnarray}
d \sigma^R - d \sigma^A & = & d\phi_{n+1} 
\left[ |M(p_1,...,p_{n+1})|^2 \theta^{cut}_{n+1}(p_1,...,p_{n+1}) \right. \nonumber \\
& & \left. - 
\sum\limits_{pairs\; i,j} \sum\limits_{k \neq i,j} {\cal D}_{ij,k}(p_1,...,p_{n+1}) \theta^{cut}_{n}(p_1,...,\tilde{p}_{ij},...,\tilde{p}_k,...,p_{n+1}) \right] . \nonumber \\
\end{eqnarray}
Both $d\sigma^R$ and $d\sigma^A$ are integrated over the same $(n+1)$ parton phase space, but it should be noted that
$d\sigma^R$ is proportional to $\theta^{cut}_{n+1}$, whereas $d\sigma^A$ is proportional to $\theta^{cut}_n$.
Here $\theta^{cut}_n$ denotes the jet-defining function for $n$-partons.
\\
\\
The subtraction term can be integrated over the one-parton phase space to yield the term
\bq
{\bf I} \otimes d\sigma^B & = & \int\limits_{1} d\sigma^A = \sum\limits_{pairs\; i,j} \sum\limits_{k \neq i,j} \int d\phi_{dipole} {\cal D}_{ij,k}.
\eq
The universal factor ${\bf I}$ still contains colour correlations, but does not depend on the unresolved parton $j$.
The term ${\bf I} \otimes d\sigma^B$ lives on the phase space of the $n$-parton configuration and has the appropriate
singularity structure to cancel the infrared divergences coming from the one-loop amplitude.
Therefore
\bq
d\sigma^V + {\bf I} \otimes d\sigma^B
\eq
is infrared finite and can easily be integrated by Monte Carlo methods.
\\
\\
In order to extend the dipole formalism to massive fermions, we have to provide three ingredients.
First, we need the correct subtraction terms for the real emission contribution.
Second, we have to integrate these subtraction terms over the dipole phase space.
The integrated terms are combined with the virtual corrections.
Third, we have to specify a mapping of the momenta, which relates the $(n+1)$-parton configuration
$p_1$,...,$p_{i}$,...,$p_j$,...$p_k$,...$p_{n+1}$ to the $n$-parton configuration $p_1$, ...,$\tilde{p}_{ij}$,...,$\tilde{p}_k$,...,$p_{n+1}$.\\
\\
In order to find appropriate dipole terms, one considers the soft and collinear
limits of the matrix element.
In the soft limit where the momentum of parton $j$ becomes soft, the $m+1$-parton matrix element behaves as
\bq
\l 1,...,m+1 | 1,...,m+1 \r = -4 \pi \mu^{2\eps} \alpha_s \l  1,...,m | {\bf J}^{\mu \dagger} {\bf J}_\mu | 1,...,m \r
\eq
with
\bq
{\bf J}^{\mu \dagger} {\bf J}_\mu | 1,...,m \r & = & \left( 2 \sum\limits_{i \neq k}  
\frac{2p_ip_k}{(2p_i p_j)  (2  p_j p_k)} {\bf T}_i \cdot {\bf T}_k
+ \sum\limits_{i} \frac{4 m_i^2}{(2p_ip_j)^2} {\bf T}_i \cdot {\bf T}_i \right) | 1,...,m \r \nonumber \\
& = & 4 \sum\limits_{i \neq k}  
\left( \frac{2p_ip_k}{(2p_i p_j)  (2 p_i p_j + 2  p_j p_k)} 
-\frac{m_i^2}{(2p_ip_j)^2} \right)
{\bf T}_i \cdot {\bf T}_k | 1,...,m \r. \nonumber \\
\eq
We used
\bq
\frac{2p_ip_k}{(2p_i p_j)  (2  p_j p_k)} & = & \frac{2p_ip_k}{(2p_i p_j)  (2 p_i p_j + 2  p_j p_k)} 
+ \frac{2p_ip_k}{(2p_j p_k)  (2 p_i p_j + 2  p_j p_k)}
\eq
and colour conservation
\bq 
\sum\limits_{i} {\bf T}_i | 1,...,m \r & = & 0.
\eq
The colour charge operators ${\bf T}_i$ for a quark, gluon and antiquark in the final state are
\bq
\mbox{quark :} & & \l ... q_i ... | T_{ij}^a | ... q_j ... \r, \nonumber \\
\mbox{gluon :} & & \l ... g^c ... | i f^{cab} | ... g^b ... \r, \nonumber \\
\mbox{antiquark :} & & \l ... \bar{q}_i ... | \left( - T_{ji}^a \right) | ... \bar{q}_j ... \r.
\eq
If the particles are massless, there is also a singularity in the collinear limit.
For final-state partons the momenta are parameterized as
\bq
\label{collinearlimit}
p_i & = & \hat{z} p + k_\perp - \frac{k_\perp^2}{\hat{z}} \frac{n}{2 p n }, \nonumber \\
p_j & = & (1-\hat{z})  p - k_\perp - \frac{k_\perp^2}{1-\hat{z}} \frac{n}{2 p n }.
\eq
Here $n$ is a massless four-vector and the transverse component $k_\perp$ satisfies
$2pk_\perp = 2n k_\perp =0$.
The collinear limits occurs for $k_\perp^2 \rightarrow 0$.
In this limit the matrix element behaves  
as
\bq
\l 1,...,m+1 | 1,...,m+1 \r & = & 4 \pi \mu^{2\eps} \alpha_s \l  1,...,m | \frac{2}{2p_ip_j} \hat{P}_{(ij),i}(\hat{z},k_\perp,\eps) | 1,...,m \r. \nonumber \\
\eq
The splitting function are given by
\bq
\l s | \hat{P}_{qq}(z,k_\perp,\eps) | s' \r & = & 
 \delta_{ss'} C_F \left[ \frac{2z}{1-z} + (1 - \rho \eps) (1-z) \right], \nonumber \\
\l s | \hat{P}_{qg}(z,k_\perp,\eps) | s' \r & = & 
 \delta_{ss'} C_F \left[ \frac{2(1-z)}{z} + (1 - \rho \eps) z \right], \nonumber \\
\l \mu | \hat{P}_{gq}(z,k_\perp,\eps) | \nu \r & = & 
 T_R \left[ -g^{\mu\nu} + 4 z (1-z) \frac{k^\mu_\perp k^\nu_\perp}{k_\perp^2} \right], \nonumber \\
\l \mu | \hat{P}_{gg}(z,k_\perp,\eps) | \nu \r & = & 
 2 C_A \left[ - g^{\mu\nu} \left( \frac{z}{1-z} + \frac{1-z}{z} \right) 
  - 2 (1-\rho \eps) z (1-z) \frac{k^\mu_\perp k^\nu_\perp}{k_\perp^2} \right].
\eq
We introduced the parameter $\rho$, which specifies the variant of dimensional
regularization:
$\rho  = 1$ for the CDR/HV schemes and $\rho=0$ for the FD scheme.
Later on we will chose the dipole terms to have the same soft and collinear
behaviour as the appropriate limit of the $(n+1)$-parton matrix element.\\
\\
If the emitting particle is in the initial state the collinear limit is defined as
\bq
p_{a} & = & p, \nonumber \\
p_i & = & (1-x) p + k_\perp - \frac{k_\perp^2}{1-x} \frac{n}{2p n}, \nonumber \\
p_{ai} & = & xp -k_\perp + \frac{k_\perp^2}{1-x} \frac{n}{2p n}.
\eq
The colour charge operators for a quark, gluon and antiquark in the initial state are
\bq
\mbox{quark :} & & \l ... \bar{q}_i ... | \left( - T_{ji}^a \right) | ... \bar{q}_j ... \r, \nonumber \\
\mbox{gluon :} & & \l ... g^c ... | i f^{cab} | ... g^b ... \r, \nonumber \\
\mbox{antiquark :} & & \l ... q_i ... | T_{ij}^a | ... q_j ... \r. 
\eq
We denoted in the amplitude an  incoming quark as an outgoing antiquark and vice versa.

\section{Calculational technique for the integration}

The integration of the dipole terms over the dipole phase space is highly non-trivial.
We first find a suitable parameterization of the phase space such that all integrals 
are of one of the following types:
\bq
\int\limits_0^1 dx x^{a-1} (1-x)^{c-a-1} & = & \frac{\Gamma(a) \Gamma(c-a)}{\Gamma(c)}, \nonumber \\
\int\limits_0^1 dx x^{a-1} (1-x)^{c-a-1} (1-x_0 x)^{-b} & = & 
 \frac{\Gamma(a) \Gamma(c-a)}{\Gamma(c)}
 {}_2F_1(a,b;c,x_0), \nonumber \\
\int\limits_0^1 dx x^{a-1} (1-x)^{c-a-1} (1-x_1 x)^{-b_1} (1-x_2 x)^{-b_2}
 & = & \frac{\Gamma(a) \Gamma(c-a)}{\Gamma(c)}
 F_1(a,b_1,b_2;c;x_1,x_2). \nonumber \\
\eq
These integrals yield the Euler-Beta function, the hypergeometric function and
the first Appell function, respectively.
The last two are then rewritten as a Taylor series in $x_0$ or $x_1$ and $x_2$, respectively.
\bq
{}_2F_1(a,b;c,x_0) & = & \sum\limits_{n=0}^\infty
 \frac{(a)_n (b)_n}{(c)_n}
 \frac{x_0^n}{n!}, \nonumber \\
F_1(a,b_1,b_2;c;x_1,x_2) & = & \sum\limits_{m_1=0}^\infty \sum\limits_{m_2=0}^\infty
 \frac{(a)_{m_1+m_2} (b_1)_{m_1} (b_2)_{m_2}}{(c)_{m_1+m_2}} 
 \frac{x_1^{m_1}}{m_1!}
 \frac{x_2^{m_2}}{m_2!}.
\eq
Here $(a)_n = \Gamma(a+n)/\Gamma(a)$ denotes the Pochhammer symbol.
Taylor expansion allows us to perform the next integral, which is of the same type as above.
After all integrals have been performed, we end up with multiple sums involving
$\Gamma$-functions depending on $\eps=(4-D)/2$.
These $\Gamma$-functions are then expanded according to
\bq
\lefteqn{
\frac{\Gamma(n+\eps)}{\Gamma(n) \Gamma(\eps)} =} & & \nonumber \\
& = & \eps \left( 1 + \eps Z_1(n-1) + \eps^2 Z_{11}(n-1) + \eps^3 Z_{111}(n-1) + .
.. 
+ \eps^{n-1} Z_{11...1}(n-1) \right), \nonumber \\
\eq
where $Z_{11...1}(n)$ are Euler-Zagier sums defined by
\bq
Z_{m_1,...,m_k}(n) & = & 
\sum\limits_{i_1 > i_2 > ... > i_k > 0}^n
\frac{1}{i_1^{m_1}}
\frac{1}{i_2^{m_2}} ...
\frac{1}{i_k^{m_k}}.
\eq
Rearranging summation indices we recognize that all sums fall in the class
of Goncharov's multiple polylogarithms \cite{Goncharov,Borwein}
\bq
\mbox{Li}_{m_k,...,m_2,m_1}(x_k,...,x_2,x_1) & = & 
 \sum\limits_{i_1 > i_2 > ... > i_k > 0}^\infty \frac{x_1^{i_1}}{i_1^{m_1}}
 \frac{x_2^{i_2}}{i_2^{m_2}}
 ...
 \frac{x_k^{i_k}}{i_k^{m_k}}.
\eq
In the case with only one massive parton it is actually sufficient to restrict oneself
to harmonic polylogarithms \cite{Vermaseren:1998uu,Remiddi:1999ew}, defined by
\bq
H_{m_1,m_2...,m_k}(x) & = & 
 \sum\limits_{i_1 > i_2 > ... > i_k > 0}^\infty \frac{x^{i_1}}{i_1^{m_1}}
 \frac{1}{i_2^{m_2}}
 ...
 \frac{1}{i_k^{m_k}}.
\eq
The additional dipole corresponding to gluon emission from a massive quark-antiquark system
can be expressed in terms of two-dimensional harmonic polylogarithms \cite{Gehrmann:2000zt}.


\section{The dipole phase space measure in D dimensions}

In this section we derive the appropriate phase space measure for the
dipoles with massive particles. Since singularities are regulated by
dimensional regularization, this has to be done in $D$ dimensions.
The phase space measure for $n$ particles in $D$ dimensions is given by \cite{Byckling}
\bq
d\phi_n(P\rightarrow p_1,...,p_n) & = & 
(2 \pi)^D \delta^D\left( P - \sum\limits_{i=1}^n p_i \right)
\prod\limits_{i=1}^n \frac{d^Dp_i}{(2\pi)^{D-1}} \theta(p_i^0) \delta(p_i^2-m_i^2) \nonumber \\
& = &
(2 \pi)^D \delta^D\left( P - \sum\limits_{i=1}^n p_i \right)
\prod\limits_{i=1}^n \frac{d^{D-1}p_i}{(2\pi)^{D-1} 2 E_i} 
\eq
with
\bq
E_i & = & \sqrt{\left(\vec{p}_i\right)^2+ m_i^2}.
\eq
The phase space measure factorizes according to
\bq
d\phi_n(P\rightarrow p_1,...,p_n) & = & 
\frac{1}{2 \pi} d \phi_{n-j+1}(P\rightarrow Q,p_{j+1},...,p_n) dQ^2
d\phi_j(Q \rightarrow p_1,...,p_j).
\eq

\subsection{Phase space measure with no initial-state particles}

We first evaluate the two particle phase space
\bq
\label{twoparticleps}
\int d\phi_2(P\rightarrow \tilde{p}_{ij}, \tilde{p}_k)
& = & 
\int \frac{d^{D-1}\tilde{p}_{ij}}{(2\pi)^{D-1} 2 E_{ij}}
\frac{d^{D-1}\tilde{p}_k}{(2\pi)^{D-1} 2 E_k}
(2 \pi)^D \delta^D(P-\tilde{p}_{ij}-\tilde{p}_k)
\eq
in the rest frame of $P$, e.g. $P=(\sqrt{P^2},\vec{0})$.
We obtain
\bq
\lefteqn{
\int d\phi_2(P\rightarrow \tilde{p}_{ij}, \tilde{p}_k)
} & & \nonumber \\ 
& = & (2\pi)^{2-D} \left(\frac{1}{2}\right)^{D-1} \left( P^2 \right)^{1-D/2}
\left( \sqrt{ \left( P^2 - m_{ij}^2 - m_k^2 \right)^2 - 4 m_{ij}^2 m_k^2 } \right)^{D-3}
\int d \Omega_{D-1},
\eq
where $\Omega_{D-1}$ parameterizes the solid angle of the $(D-1)$ spatial components of 
$\tilde{p}_k$ in $(D-1)$ (spatial) dimensions.
With this convention we have
\bq
\int d\Omega_D & = & \frac{2 \pi^{D/2}}{\Gamma\left( D/2 \right)}.
\eq
We then evaluate the three particle phase space
\bq
\lefteqn{
\int d\phi_3(P\rightarrow p_i, p_j, p_k) } & & \nonumber \\
& = & 
\int \frac{d^{D-1}p_i}{(2\pi)^{D-1} 2 E_i}
\frac{d^{D-1}p_j}{(2\pi)^{D-1} 2 E_j}
\frac{d^{D-1}p_k}{(2\pi)^{D-1} 2 E_k}
(2 \pi)^D \delta^D(P-p_i-p_j-p_k)
\eq
in the rest frame of $P$, e.g. $P=(\sqrt{P^2},\vec{0})$.
We shall orient our frame such that the solid angle of the spatial components
of $p_k$ coincides with the solid angle of the spatial components of 
$\tilde{p}_k$ in (\ref{twoparticleps}).
It will be convenient to parameterize the spatial components of $p_i$
with spherical coordinates, using $\vec{p}_k$ as polar axis, e.g
\bq
d^{D-1}p_i & = & \left| \vec{p}_i \right|^{D-2} d \left| \vec{p}_i \right|
d \theta_1 \sin^{D-3} \theta_1 d \Omega_{D-2}^{(i)}.
\eq
We therefore have
\bq
2 \vec{p}_i \vec{p}_k & = & 2 \left| \vec{p}_i \right| \left| \vec{p}_k \right|
\cos \theta_1.
\eq
Finally we obtain the three-particle phase space as the product of a two-particle phase
space and a dipole phase space:
\bq
\int d\phi_3(P\rightarrow p_i, p_j, p_k)
& = & 
\int d\phi_2(P\rightarrow \tilde{p}_{ij}, \tilde{p}_k)
d\phi_{dipole}(2p_ip_j,2p_jp_k,2p_ip_k,\Omega_{D-2}^{(i)}),
\eq
where
\bq
\lefteqn{
d\phi_{dipole}(2p_ip_j,2p_jp_k,2p_ip_k)
=  
(2\pi)^{1-D} 
\frac{2\pi^{D/2-1}}{\Gamma\left(\frac{D}{2}-1\right)}
\frac{1}{4}
\left( \left( P^2 - m_i^2 - m_k^2 \right)^2 - 4 m_i^2 m_k^2 \right)^{\frac{3-D}{2}} } & & 
\nonumber \\
& & 
\int d(2p_ip_j) d(2p_jp_k) d(2p_ip_k) \delta\left(P^2-m_i^2-m_j^2-m_k^2-2p_ip_j -2 p_jp_k -2 p_ip_k\right)
\nonumber \\
& & 
\left(  -P^2 \lambda\left( |\vec{p}_i|^2,|\vec{p}_j|^2,|\vec{p}_k|^2 \right) \right)^{\frac{D-4}{2}}
\theta\left( - \lambda\left( |\vec{p}_i|^2,|\vec{p}_j|^2,|\vec{p}_k|^2 \right) \right).
\eq
Note that we already performed the angular integration over
$d\Omega_{D-2}^{(i)}$.
The triangle function $\lambda$ is defined by
\bq
\lambda(x,y,z) & = & x^2+y^2+z^2-2xy-2yz-2zx.
\eq
We further have
\bq
|\vec{p}_i|^2 & = & \frac{1}{4P^2}\left( P^2+m_i^2-m_k^2-m_j^2-2p_jp_k \right)^2 - m_i^2,
\nonumber \\
|\vec{p}_k|^2 & = & \frac{1}{4P^2}\left( P^2-m_i^2+m_k^2-m_j^2-2p_ip_j \right)^2 - m_k^2,
\nonumber \\
|\vec{p}_j|^2 & = & \frac{1}{4P^2}\left( P^2-m_i^2-m_k^2+m_j^2-2p_ip_k \right)^2 - m_j^2 
\eq
so that
\bq
\lefteqn{
-P^2 \lambda\left( |\vec{p}_i|^2,|\vec{p}_j|^2,|\vec{p}_k|^2 \right) } & & \nonumber \\
& = & 2p_ip_k 2p_ip_j 2 p_jp_k 
- m_i^2 \left( 2p_jp_k \right)^2
- m_k^2 \left( 2p_ip_j \right)^2
- m_j^2 \left( 2p_ip_k \right)^2
+ 4 m_i^2 m_j^2 m_k^2.
\eq
To find a suitable parametrization of the dipole phase space, we treat the various cases
separately. 
We consider first the case where two particles are massless $m_i=m_j=0$.
With $D=4-2\eps$ the dipole phase space becomes
\bq
d\phi_{dipole} 
& = & 
\frac{(4\pi)^{\eps-2}}{\Gamma\left(1-\eps\right)}  
\left( P^2 \right)^{1-\eps} \left(u_0\right)^{2-2\eps}
\nonumber \\
& & 
\cdot 
\int\limits_0^1 du u^{1-2\eps} (1-u)^{1-2\eps} \left( 1 - u_0 u \right)^{-1+\eps}
\int\limits_0^1 dv v^{-\eps} (1-v)^{-\eps},
\eq
where
\bq
u_0 & = & \frac{P^2-m_k^2}{P^2}, \nonumber \\
u & = & \frac{2p_ip_j+2p_jp_k}{2p_ip_k+2p_ip_j+2p_jp_k} = \frac{2p_ip_j+2p_jp_k}{P^2-m_k^2}, \nonumber \\
v & = & \frac{2p_ip_j \left( 2p_ip_k + m_k^2 \right)}{2p_ip_k \left( 2p_ip_j+2p_jp_k\right)}
= \frac{2p_ip_j \left( 2p_ip_k + m_k^2 \right)}{2p_ip_k \left( P^2 -m_k^2 - 2 p_ip_k \right)}.
\eq
In the case where one mass vanishes ($m_j=0$) and the other two are equal
($m_i=m_k=m$) we obtain
\bq
d\phi_{dipole} 
& = & 
\frac{(4\pi)^{\eps-2}}{\Gamma\left(1-\eps\right)}  
\left( P^2 \right)^{1-\eps} \left(r_0\right)^{2-2\eps} 2^{2\eps-1} 
\nonumber \\
& & 
\cdot
\int\limits_0^1 dr r^{1-2\eps} (1-r)^{-\eps+1/2} \left( 1 - r_0 r \right)^{-1/2}
\int\limits_{-1}^1 ds \left( 1 - s^2 \right)^{-\eps},
\eq
with
\bq
r_0 & = & 1 -4 \frac{m^2}{P^2}, \nonumber \\
2 p_i p_j & = & \frac{1}{2} P^2 r_0 r \left( 1 - s \sqrt{ \frac{r_0(1-r)}{1-r_0r} } \right),
\nonumber \\
2 p_j p_k & = & \frac{1}{2} P^2 r_0 r \left( 1 + s \sqrt{ \frac{r_0(1-r)}{1-r_0r} } \right).
\eq

\subsection{Phase space involving initial state particles}

If initial state particles are involved, we obtain the following convolution:
\bq
\label{initial_1}
d\phi(p_a+p_b \rightarrow K + p_k + p_i ) & = & \int\limits_0^1 dx d\phi(\tilde{p}_a+p_b
\rightarrow K + \tilde{p}_k) d\phi_{dipole}
\eq
with 
\bq
\tilde{p}_a & = & x p_a.
\eq
In the case $m_a=m_i=0$ we obtain
\bq
\label{initial_2}
\lefteqn{
d\phi_{dipole} } & & \nonumber \\
& = & 
\frac{(4 \pi)^{\eps-2}}{\Gamma(1-\eps)} \left(-P^2\right)^{1-\eps}
x_0^{\eps-1} x^{-1+\eps} (1-x)^{1-2\eps} (1-x_0x)^{-1+\eps}
\int\limits_0^1 dw w^{-\eps} (1-w)^{-\eps}
\eq
with $P=p_k+p_i-p_a$ and
\bq
x_0 & = & \frac{-P^2}{m_k^2-P^2}, \nonumber \\
x & = & \frac{2p_ap_i+2p_ap_k-2p_ip_k}{2p_ap_i+2p_ap_k}, \nonumber \\
w & = & \frac{2p_ap_i \left(2p_ip_k +m_k^2 \right)}{2p_ip_k \left(2p_ap_i+2p_ap_k\right)}.
\eq
Eq. (\ref{initial_1}) and eq. (\ref{initial_2}) are derived as follows: From the factorization of phase space we have
\bq
d\phi(p_a+p_b \rightarrow K + p_k + p_i ) & = & \frac{1}{2\pi} d\phi(p_a+p_b \rightarrow K+Q) dm_Q^2 d\phi(Q \rightarrow p_i+p_k).
\eq
We then derive 
\bq
d\phi(p_a+p_b \rightarrow K+Q) & = & x d\phi(\tilde{p}_a+p_b \rightarrow K+\tilde{p}_k).
\eq
This equation can be obtained by writing out the explicit expressions for the phase space measures in the rest frame of
$p_a+p_b$ and $\tilde{p}_a+p_b$, respectively. These two frames are related by a boost. Further, since $\tilde{p}_a = x p_a$, the boost does
not affect the transverse components.
Next, one writes out the parameterization for $d\phi(Q \rightarrow p_i+p_k)$ in terms of solid angles of particle $p_k$, singles out one angle
$\theta_1 = \angle(p_k,p_a)$ and replaces the integrations over $dm_Q^2 d\cos\theta_1$ by integrations over $dx dw$.

\section{The dipole terms for final-state singularities with no initial-state partons}

In this section we give the dipole factors corresponding to the case where all relevant
partons are in the final state.
Initial-state partons are considered in the next section.
We distinguish the cases of (i) a massless emitter and massive spectator,
(ii) a massive emitter and a massless spectator and (iii) a massive emitter and
a massive spectator of equal mass.
We follow closely the notation of Catani and Seymour \cite{Catani:1997vz}.\\

The generic form of the dipole terms is given by
\bq
{\cal D}_{ij,k} & = &
- \frac{1}{2 p_i \cdot p_j}
\langle 1, ..., \tilde{(ij)},...,\tilde{k},...|
\frac{{\bf T}_k \cdot {\bf T}_{ij}}{{\bf T}^2_{ij}} V_{ij,k} |
1,...,\tilde{(ij)},...,\tilde{k},... \rangle.
\eq
The explicit forms of the functions $V_{ij,k}$ are given below.

\subsection{Massless final emitter, massive final spectator}

We consider first the case of massless emitter (particle $i$) and a massive
spectator (particle $k$), both in the final state.
The variables $y$ and $z$ of Catani and Seymour are given by
\bq
y = \frac{2p_ip_j}{2p_ip_j+2p_jp_k+2p_ip_k}, & & 
z = \frac{2p_ip_k}{2p_ip_k+2p_jp_k}
\eq
and related to the variables $u$ and $v$ as follows:
\bq
y & = & u_0 v \frac{u(1-u)}{1-u_0u}, \;\;\; u_0 = 1 -\frac{m_k^2}{(p_i+p_j+p_k)^2} \nonumber \\
z (1-y) & = & 1-u.
\eq
In the collinear limit eq.(\ref{collinearlimit}) we have
\bq y \rightarrow 0, & & z \rightarrow \hat{z} 
\eq
and in the soft limit $p_j \rightarrow 0$
\bq
y \rightarrow 0, & & z \rightarrow 1.
\eq
The subtraction term for the splitting $g\rightarrow g g$ has in addition to match the soft limit
$p_i \rightarrow 0$, corresponding to $y \rightarrow 0$, $z \rightarrow 0$.\\
  
As subtraction terms we use
\bq
\l s | V_{q_ig_j,k} | s' \r & = & 8 \pi \mu^{2\eps} \alpha_s C_F \delta_{ss'}
\left[ \frac{2z(1-y)}{1-z(1-y)} + (1-\rho \eps) (1-z(1-y)) \right], \nonumber \\
\l \mu | V_{q_i \bar{q}_j,k} | \nu \r & = & 8 \pi \mu^{2\eps} \alpha_s T_R
\left[ -g^{\mu\nu} - \frac{4}{2p_ip_j} S^{\mu\nu} \right], \nonumber \\
\l \mu | V_{g_ig_j,k} | \nu \r & = & 16 \pi \mu^{2\eps} \alpha_s C_A
\left[ -g^{\mu\nu} \left( \frac{z(1-y)}{1-z(1-y)}
                         +\frac{(1-z)(1-y)}{1-(1-z)(1-y)} 
                   \right) 
 \right. \nonumber \\
 & & \left.
 + (1-\rho \eps) \frac{2}{2p_ip_j} S^{\mu\nu} \right],
\eq
where the spin correlation tensor is given by
\bq
S^{\mu\nu} & = & (z(1-y) p_i^\mu - (1-z(1-y)) p_j^\mu) ( z(1-y)p_i^\nu - (1-z(1-y)) p_j^\nu).
\eq
The momenta are mapped as follows:
\bq
\tilde{p}_i & = & a p_i + b p_j + c p_k, \nonumber \\
\tilde{p}_k & = & (1-a) p_i + (1-b) p_j + (1-c) p_k,
\eq
with
\bq
c & = & \frac{y u_0}{(2u-1-y(1-u))^2u_0+4u(1-u)y} \left( 
2u(1-u) - \frac{N}{\sqrt{1-v}} \right), \nonumber \\
b & = & \frac{1}{N} \left\{ u + \frac{c}{y u_0} \left[ -2yu - u_0 \left( (1-2u-u^2)y+1-3u+2u^2 \right) \right] \right\}, \nonumber \\
a & = & \frac{1}{N} \left\{ 1-u - \frac{c}{y u_0} \left[ 2y(1-u) - u_0 \left( (1-u) y^2 + (1-u+u^2) y + u -2 u^2 \right) \right] \right\}, \nonumber \\
N & = & u^2 + (1-u)^2 + (1-u) y.
\eq
This mapping satisfies momentum conservation
\bq
\tilde{p}_i + \tilde{p}_k & = & p_i + p_j + p_k,
\eq
and the on-shell conditions
\bq
\tilde{p}_i^2 = 0, & & \tilde{p}_k^2 = m_k^2.
\eq
In the limit $y \rightarrow 0$ we have
\bq
\lim\limits_{y \rightarrow 0} \tilde{p}_i = p_i+p_j, 
& & 
\lim\limits_{y \rightarrow 0} \tilde{p}_k = p_k.
\eq
In addition we have
\bq
S_{\mu\nu} \tilde{p}_i^\nu & = & 0,
\eq
e.g. $\tilde{p}_i$ is orthogonal to the spin correlation tensor.
The integral over the spin correlation can be written as
\bq
\int d\phi_{dipole} \frac{-2}{(2p_ip_j)^2} S^{\mu\nu} & = &
C_{21} \tilde{p}_i^\mu \tilde{p}_i^\nu + C_{22} \tilde{p}_k^\mu \tilde{p}_k^\nu
+ C_{23} \left( \tilde{p}_i^\mu \tilde{p}_k^\nu + \tilde{p}_k^\mu \tilde{p}_i^\nu \right)
+ C_{24} g^{\mu\nu}.
\eq
Using $S_{\mu\nu} \tilde{p}_i^\nu = 0$
and $\tilde{p}_i^2=0$
this reduces to 
\bq
C_{21} \tilde{p}_i^\mu \tilde{p}_i^\nu 
- C_{24} \left( -g^{\mu\nu} + 2 \frac{   \tilde{p}_i^\mu \tilde{p}_k^\nu 
                                       + \tilde{p}_k^\mu \tilde{p}_i^\nu}
                                     {2 \tilde{p}_i \tilde{p}_k }
         \right).
\eq
Due to gauge invariance only the term $-C_{24} (-g^{\mu\nu})$ will give a non-vanishing
contribution.
$C_{24}$ is obtained by contracting with $g_{\mu\nu}$:
\bq
C_{24} & = & \frac{1}{2(1-\rho\eps)} \int d\phi_{dipole} \frac{-2}{(2p_ip_j)^2} 
  g_{\mu\nu} S^{\mu\nu}
\eq
Integration of the subtraction terms yields:
\bq
{\cal V}_{q_ig_j,k}  & = & \int d\phi_{dipole} \frac{1}{2p_ip_j} V_{q_ig_j,k} 
 = \frac{\alpha_s}{2\pi} \frac{1}{\Gamma(1-\eps)} \left( \frac{4 \pi \mu^2 P^2}{(P^2-m_k^2)^2} \right)^\eps {\cal V}_{qg}(u_0,\eps), \nonumber \\
\lefteqn{
{\cal V}_{qg}(u_0,\eps) } & & \nonumber \\
& = & C_F \delta_{ss'} 
\int\limits_0^1 du \int\limits_0^1 dv 
 u^{-2\eps} (1-u)^{-2\eps} (1-u_0u)^\eps 
 v^{-1-\eps} (1-v)^{-\eps} \left[ 2 \frac{1-u}{u} + ( 1 - \rho \eps ) u \right] \nonumber \\
& = & C_F \delta_{ss'} \frac{\Gamma(-\eps) \Gamma(1-\eps)}{\Gamma(1-2\eps)} \Gamma(2-2\eps) 
\left[ 2 \frac{\Gamma(-2\eps)}{\Gamma(2-4\eps)} {}_2F_1(-\eps,-2\eps;2-4\eps;u_0) \right. \nonumber \\
& & \left. + (1-\rho \eps) \frac{\Gamma(1-2\eps)}{\Gamma(3-4\eps)} {}_2F_1(-\eps,2-2\eps;3-4\eps;u_0) \right] \nonumber \\
& = & C_F \delta_{ss'} \left[ \frac{1}{\eps^2} + \frac{3}{2\eps} + \frac{17}{4} 
+ \frac{1}{2} \rho - \frac{5}{6} \pi^2 + \frac{1}{2u_0}
 + \frac{1}{2} \frac{(1-u_0)(1-3u_0)}{u_0^2} \ln(1-u_0) \right. \nonumber \\
 & & \left. + 2 \mbox{Li}_2(u_0) \right] 
 + O(\eps), \nonumber \\
{\cal V}_{q_i\bar{q}_j,k}  & = & \int d\phi_{dipole} \frac{1}{2p_ip_j} V_{q_i\bar{q}_j,k} 
 = \frac{\alpha_s}{2\pi} \frac{1}{\Gamma(1-\eps)} \left( \frac{4 \pi \mu^2 P^2}{(P^2-m_k^2)^2} \right)^\eps {\cal V}_{q\bar{q}}(u_0,\eps), \nonumber \\
\lefteqn{
{\cal V}_{q\bar{q}}(u_0,\eps) } & & \nonumber \\
& = &  
\int\limits_0^1 du \int\limits_0^1 dv 
 u^{-2\eps} (1-u)^{-2\eps} (1-u_0u)^\eps 
 v^{-1-\eps} (1-v)^{-\eps} 
 T_R \left[ -g^{\mu\nu} -\frac{4}{2p_ip_j} S^{\mu\nu}\right] \nonumber \\
& = & T_R (-g^{\mu\nu}) \frac{\Gamma(-\eps)\Gamma(1-\eps)}{\Gamma(1-2\eps)} 
 \left[ 
  \frac{\Gamma(1-2\eps)^2}{\Gamma(2-4\eps)} {}_2F_1(-\eps,1-2\eps,2-4\eps,u_0)
 \right. \nonumber \\
 & & \left.
   - \frac{2}{1-\rho\eps} \frac{\Gamma(2-2\eps)^2}{\Gamma(4-4\eps)} 
     {}_2F_1(-\eps,2-2\eps,4-4\eps,u_0)
 \right] + \; \mbox{gauge terms} \nonumber \\
& = & 
 T_R (-g^{\mu\nu} ) \left[ -\frac{2}{3\eps} -\frac{13}{6} + \frac{1}{3} \rho 
      + \frac{2}{3} \frac{(1-u_0)}{u_0^2} \right. \nonumber \\
 & & \left. + \frac{1}{3} \frac{(1-u_0)}{u_0^3} 
          \left( 2u_0^2 -u_0+2\right) \ln(1-u_0) \right] 
 + \; \mbox{gauge terms} \; + O(\eps), \nonumber \\
{\cal V}_{g_ig_j,k}  & = & \int d\phi_{dipole} \frac{1}{2p_ip_j} V_{g_ig_j,k} 
 = \frac{\alpha_s}{2\pi} \frac{1}{\Gamma(1-\eps)} \left( \frac{4 \pi \mu^2 P^2}{(P^2-m_k^2)^2} \right)^\eps {\cal V}_{gg}(u_0,\eps), \nonumber \\
\lefteqn{
{\cal V}_{gg}(u_0,\eps) } & & \nonumber \\
& = & 2 C_A  
\int\limits_0^1 du \int\limits_0^1 dv 
 u^{-2\eps} (1-u)^{-2\eps} (1-u_0u)^\eps 
 v^{-1-\eps} (1-v)^{-\eps} \nonumber \\
 & & \cdot 
 \left[ -g^{\mu\nu} \left( \frac{1-u}{u} + \frac{u}{1-u} 
                            - \frac{u}{(1-u)} \frac{u_0 v}{(1-u_0u(1-v))}
                    \right)
 + (1-\rho\eps) \frac{2}{2p_ip_j} S^{\mu\nu}\right] \nonumber \\
& = & 2 C_A (-g^{\mu\nu}) \frac{\Gamma(-\eps)\Gamma(1-\eps)}{\Gamma(1-2\eps)} 
 \left[ 
  \frac{\Gamma(-2\eps) \Gamma(2-2\eps)}{\Gamma(2-4\eps)} \left( 
    {}_2F_1(-\eps,-2\eps,2-4\eps,u_0) 
    \right. \right. \nonumber \\ & & \left. \left. 
     + {}_2F_1(-\eps,2-2\eps,2-4\eps,u_0) \right)
   + \frac{\Gamma(2-2\eps)^2}{\Gamma(4-4\eps)} 
     {}_2F_1(-\eps,2-2\eps,4-4\eps,u_0)
 \right] \nonumber \\
 & & 
  + 2 C_A (-g^{\mu\nu}) (-u_0) \frac{\Gamma(1-\eps)\Gamma(-2\eps)}{\Gamma(-\eps)} 
    \sum\limits_{n=0}^\infty \sum\limits_{m=0}^\infty
    \frac{\Gamma(n+1-\eps)}{\Gamma(n+2-2\eps)} \frac{\Gamma(m+n+2-2\eps)}{\Gamma(m+n+2-4\eps)} 
    \nonumber \\
 & &
    \frac{\Gamma(m-\eps)}{m!} u_0^{n+m}
    + \; \mbox{gauge terms} \nonumber \\
& = & 
 2 C_A (-g^{\mu\nu} ) \left[ 
  \frac{1}{\eps^2} 
  + \frac{11}{6} \frac{1}{\eps}
  -\frac{5}{6} \pi^2 + \frac{67}{12} 
  - \frac{1}{3} \frac{(1-u_0)}{u_0^2}
  + 2\; \mbox{Li}_2(u_0)
\right. \nonumber \\ & & \left. 
                            - \frac{1}{6} \left( \frac{(1-u_0)}{u_0^3} (2-u_0) 
                                                  + 11 \frac{(1-u_0)}{u_0} 
                                           \right) \ln(1-u_0)
                      \right] \nonumber \\
 & & + \; \mbox{gauge terms} \; + O(\eps). 
\eq

\subsection{Massive final emitter, massless final spectator}

We now consider the case of a massive emitter (particle $k$) and a massless spectator (particle $i$).
It is sufficient to consider the case where a heavy quark (or antiquark) emits a gluon.
We now have
\bq
y = \frac{2p_jp_k}{2p_ip_j+2p_jp_k+2p_ip_k}, & & 
z = \frac{2p_ip_k}{2p_ip_k+2p_ip_j}
\eq
and
\bq
y & = & u \left( 1 -\frac{(1-u) u_0}{1-u_0 u}  v \right), \;\;\; u_0 = 1 -\frac{m_k^2}{(p_i+p_j+p_k)^2}\nonumber \\
z (1-y) & = & 1-u.
\eq
The momenta are mapped as follows:
\bq
\mbox{emitter :} & & \tilde{p}_k = p_k + p_j - \frac{y}{1-y} p_i, \nonumber \\
\mbox{spectator :} & & \tilde{p}_i = \frac{1}{1-y} p_i.
\eq
Since there is no collinear singularity, we just have to match the part of the soft singularity which
corresponds to this dipole factor:
\bq
\l s | V_{q_kg_j,i} | s' \r & = & 8 \pi \mu^{2\eps} \alpha_s C_F \delta_{ss'}
\left[ \frac{2z(1-y)}{1-z(1-y)}  -2 \left( \frac{1-u_0}{u_0} \right) \frac{1}{y} \right] 
\eq
With
\bq
 v_0 & = & \frac{u_0 (1-u)}{1-u_0 u}
\eq
we obtain for the integral over the dipole phase space
\bq
{\cal V}_{q_kg_j,i} & = & \int d\phi_{dipole} \frac{1}{2p_jp_k} V_{q_kg_j,i} = \frac{\alpha_s}{2\pi} \frac{1}{\Gamma(1-\eps)}
\left( \frac{4\pi \mu^2 P^2}{(P^2-m_k)^2} \right)^\eps {\cal V}_{Qg}(u_0,\eps),
\eq
\bq
{\cal V}_{Qg}(u_0,\eps) & = &
C_F \delta_{ss'} \cdot 2 \int\limits_0^1 du \int\limits_0^1 dv
 u^{-2\eps-1} (1-u)^{1-2\eps} (1-u_0 u)^{\eps-1}
 v^{-\eps} (1-v)^{-\eps} (1-v_0 v)^{-1} \nonumber \\
 & &
  \left[ u_0 (1-u) - (1-u_0) (1 -v_0 v)^{-1} \right] \nonumber \\ 
& = & 
C_F \delta_{ss'} \cdot 2 \Gamma(1-\eps) 
\sum\limits_{n=0}^\infty \sum\limits_{m=0}^\infty
\frac{\Gamma(m-2\eps) \Gamma(n+m+1-\eps)}{\Gamma(n+m+2-4\eps)} \nonumber \\
& & \cdot 
\left( \frac{n+2-2\eps}{n+m+2-4\eps} u_0 - (n+1) (1-u_0) \right) \frac{u_0^{n+m}}{m!} \nonumber \\
& = & C_F \delta_{ss'} \left[ \frac{1}{\eps} \left(1+ \ln(1-u_0) \right)
+4 + \ln(1-u_0) -4 \mbox{Li}_2(u_0) - \frac{1}{2} \ln^2(1-u_0) \right] \nonumber \\
& & + O(\eps).
\eq
Here we used 
\bq
\sum\limits_{i_1=2}^\infty \sum\limits_{i_2=1}^{i_1-1} 
 \frac{u_0^{i_1}}{i_1} \frac{1}{i_2} 
 & = & \mbox{Li}_{1,1}(1,u_0) = H_{11}(u_0) = \frac{1}{2} \ln^2(1-u_0),
\nonumber \\
 \sum\limits_{i_1=2}^\infty \sum\limits_{i_2=1}^{i_1-1}
 u_0^{i_1} \frac{1}{i_2} 
 & = & \mbox{Li}_{1,0}(1,u_0) = -\frac{u_0}{1-u_0} \ln(1-u_0).
\eq  

\subsection{Massive final emitter and massive final spectator of equal mass}

We now consider the case of a massive emitter (particle $i$) and a massive spectator (particle $k$) of equal mass ($m_i=m_k=m$).
It is sufficient to consider the case where a heavy quark (or antiquark) emits a gluon.
Since there is no collinear singularity, we just have to match the part of the soft singularity which
corresponds to this dipole factor:
\bq
\lefteqn{
\l s | V_{q_ig_j,k} | s' \r } & & \nonumber \\
& = & 8 \pi \mu^{2\eps} \alpha_s C_F \delta_{ss'}
\sqrt{ \frac{(2p_ip_j+2p_jp_k+2p_ip_k)^2-(2m^2)^2}{(2p_ip_k)^2-(2m^2)^2} }
\left[ 2 \frac{2p_ip_k}{2p_ip_j+2p_jp_k} - 2 \frac{m^2}{2p_ip_j} \right] \nonumber \\
 & = & 8 \pi \mu^{2\eps} \alpha_s C_F \delta_{ss'} \frac{1}{r_0 r \sqrt{(1-r)(1-r_0r)}}
\left[ 2 (1-r_0 r) - (1-r_0) - \frac{1-r_0}{1 - s_0 s}
\right], \nonumber \\
\eq
where 
\bq 
r & = & \frac{2p_ip_j+2p_jp_k}{2p_ip_j+2p_jp_k+2p_ip_k-2m^2}, \nonumber \\
s & = & - \frac{1}{s_0} \frac{2p_ip_j-2p_jp_k}{2p_ip_j+2p_jp_k}, \nonumber \\
s_0 & = & \sqrt{ \frac{r_0(1-r)}{1-r_0r} }, 
\;\;\; r_0 = 1 - 
\frac{4 m^2}{(p_i+p_j+p_k)^2}.
\eq
The singularity occurs for $r \rightarrow 0$. In this limit, the expression in the square
root tends to $1$. The inclusion of the square root term facilitates the analytic integration
of the dipole term.
The momenta are mapped as follows:
\bq
\tilde{p}_i & = & \frac{1}{2} P - \sqrt{ \frac{r_0}{y_0^2-(1-r_0)}} \left( p_k - \frac{y_0}{2} P \right),
 \nonumber \\
\tilde{p}_k & = & \frac{1}{2} P + \sqrt{ \frac{r_0}{y_0^2-(1-r_0)}} \left( p_k - \frac{y_0}{2} P \right),
\eq
with $P=p_i+p_j+p_k$ and
\bq
y_0 & = & \frac{2p_ip_k+2p_jp_k+2m^2}{P^2}.
\eq 
This mapping satisfies the on-shell conditions $\tilde{p}_i^2 = \tilde{p}_k^2 = m^2$.
In the soft limit $y_0$ tends to $1$ and the correct asymptotic behaviour of the mapping
is easily verified. Integration yields:
\bq
{\cal V}_{q_ig_j,k} & = & \int d\phi_{dipole} \frac{1}{2p_ip_j} V_{q_ig_j,k} = \frac{\alpha_s}{2\pi} \frac{1}{\Gamma(1-\eps)}
\left( \frac{4\pi \mu^2 }{P^2} \right)^\eps {\cal V}_{QQ}(r_0,\eps),
\eq
\bq
{\cal V}_{QQ}(r_0,\eps) & = & 
 r_0^{-2\eps} 2^{2\eps} \int\limits_0^1 dr r^{-2\eps-1} (1-r)^{-\eps} (1-r_0r)^{-1}
 \int\limits_{-1}^1 ds (1-s^2)^{-\eps} \nonumber \\
 & & 
 C_F \delta_{ss'} \left[
 \left( 2(1-r_0r) -(1-r_0) \right) (1-s_0 s)^{-1} -(1-r_0) (1-s_0 s)^{-2} \right] \nonumber \\
& = & 
 C_F \delta_{ss'} \frac{1}{1-\eps} 2^{2\eps} r_0^{-2\eps} 
 \sum\limits_{i=0}^\infty
 \sum\limits_{j=0}^\infty
 \sum\limits_{k=0}^\infty
  (-1)^i \left( 1 + (-1)^j \right) r_0^{k+j/2} \nonumber \\
 & &
  \frac{\Gamma(1+i+j) \Gamma(2-\eps)}{\Gamma(2+i+j-\eps)}
  \frac{\Gamma(i+\eps)}{\Gamma(\eps) i!}
  \frac{ \Gamma(k+j/2) \Gamma(k-2\eps) \Gamma(1+j/2-\eps)}{\Gamma(j/2) 
         \Gamma(k+j/2+1-3\eps) k!} \nonumber \\
 & &
  \left[ 2 - (1-r_0) (2+j) \frac{j+2k}{j} \right]
 \nonumber \\
& = & 
C_F \delta_{ss'} \left\{
  \frac{1}{\eps} \left( 1 - \frac{1}{2} \frac{1+r_0}{\sqrt{r_0}} 
                            \ln \frac{1+\sqrt{r_0}}{1-\sqrt{r_0}} \right) 
  - 2 \ln r_0 
 \right. \nonumber \\
 & &
 - \ln^2\left(\frac{1+\sqrt{r_0}}{1-\sqrt{r_0}}\right) +\frac{1}{\sqrt{r_0}}\ln\left(\frac{1+\sqrt{r_0}}{1-\sqrt{r_0}}\right)
 \nonumber \\
 & & \left.
 - \frac{1+r_0}{2\sqrt{r_0}} 
   \left( \mbox{Li}_2\left(\sqrt{r_0}\right) - \mbox{Li}_2\left(-\sqrt{r_0}\right)
          +2\;\mbox{Li}_2\left(\frac{1+\sqrt{r_0}}{2}\right)-2\;\mbox{Li}_2\left(\frac{1-\sqrt{r_0}}{2}\right) 
 \right. \right. \nonumber \\
 & & \left. \left.
          +\mbox{Li}_2\left(\frac{\sqrt{r_0}-1}{2\sqrt{r_0}}\right)
          -\mbox{Li}_2\left(\frac{\sqrt{r_0}-1}{\sqrt{r_0}}\right)
          +\mbox{Li}_2\left(\frac{1}{1+\sqrt{r_0}}\right)
          -\mbox{Li}_2\left(\frac{1-\sqrt{r_0}}{1+\sqrt{r_0}}\right)
 \right. \right. \nonumber \\
 & & \left. \left.
          -2\ln r_0 \ln\left(\frac{1+\sqrt{r_0}}{1-\sqrt{r_0}}\right)
          +\ln2 \ln \frac{\sqrt{r_0}}{1+\sqrt{r_0}}
          +\frac{1}{2} \ln^2 2
 \right. \right. \nonumber \\
 & & \left. \left.
          + \ln(1-\sqrt{r_0}) \ln \left( \frac{1+\sqrt{r_0}}{\sqrt{r_0}} \right)
          + \frac{1}{2} \ln^2(1+\sqrt{r_0}) 
          - \frac{1}{2} \ln^2(1-\sqrt{r_0})
   \right) 
 \right\}
 \nonumber \\
& & + O(\eps).
\eq
Here we used
\bq
\lefteqn{
\sum\limits_{i_1=2}^\infty \sum\limits_{i_2=1}^{i_1-1} 
 \frac{x_1^{i_1}}{i_1} \frac{x_2^{i_2}}{i_2} 
 = \mbox{Li}_{1,1}(x_2,x_1) } & & \nonumber \\ 
 & = &
 \ln(1-x_1) \ln(1-x_2) + \mbox{Li}_2\left(-\frac{x_2}{1-x_2} \right)
  -\mbox{Li}_2\left(- \frac{x_2(1-x_1)}{1-x_2} \right) \nonumber \\
 & = & 
 -\frac{1}{2} \ln^2\left(1-x_1x_2\right) +\ln\left(x_2(1-x_1)\right) \ln(1-x_1x_2)
  -\mbox{Li}_2\left(\frac{1-x_2}{1-x_1x_2}\right) + \mbox{Li}_2\left(1-x_2\right), \nonumber \\
\lefteqn{
\sum\limits_{i_1=2}^\infty \sum\limits_{i_2=1}^{i_1-1} 
 x_1^{i_1} \frac{x_2^{i_2}}{i_2} 
 = \mbox{Li}_{1,0}(x_2,x_1) = - \frac{x_1}{1-x_1} \ln(1-x_1x_2), } & &\nonumber \\
\lefteqn{
\sum\limits_{i_1=2}^\infty \sum\limits_{i_2=1}^{i_1-1} 
 \frac{x_1^{i_1}}{i_1} x_2^{i_2} 
 = \mbox{Li}_{0,1}(x_2,x_1) = 
 \frac{1}{1-x_2} \ln(1-x_1x_2) - \frac{x_2}{1-x_2} \ln(1-x_1). } & &
\eq

\section{The dipole terms with initial-state partons}

In this section we consider initial-state partons.
We distinguish the cases of (i) a massless emitter in the initial state and massive spectator
in the final state,
(ii) a massive emitter in the final state and a massless spectator in the initial state.
The generic form of the dipole terms for case (i) is given by
\bq
{\cal D}^{ai}_{k} & = &
- \frac{1}{2 p_a \cdot p_i} \frac{1}{x}
\langle 1, ..., \tilde{(ai)},...,\tilde{k},...|
\frac{{\bf T}_k \cdot {\bf T}_{ai}}{{\bf T}^2_{ai}} V^{ai}_{k} |
1,...,\tilde{(ai)},...,\tilde{k},... \rangle.
\eq
In the case (ii) we have
\bq
{\cal D}^{a}_{ki} & = &
- \frac{1}{2 p_i \cdot p_k} \frac{1}{x}
\langle 1, ..., \tilde{(ki)},...,\tilde{a},...|
\frac{{\bf T}_a \cdot {\bf T}_{ki}}{{\bf T}^2_{ki}} V^{a}_{ki} |
1,...,\tilde{(ki)},...,\tilde{a},... \rangle.
\eq
The explicit forms of the functions $V^{ai}_{k}$ and $V^{a}_{ki}$
are given below.

\subsection{Massless initial emitter, massive final spectator}

We now consider the case of a massless emitter (particle $a$) in the initial state
and a massive spectator (particle $k$) in the final state.
The variables $u$ and $x$ of Catani and Seymour are given by
\bq
u = \frac{2p_ip_a}{2p_ip_a+2p_kp_a}, & & x = \frac{2p_ip_a+2p_kp_a-2p_ip_k}{2p_ip_a+2p_kp_a}.
\eq
The variables $u$ and $w$ are related by
\bq
u & = & \frac{1-x}{1-x_0x} w, \;\;\; x_0 = 1- \frac{m_k^2}{m_k^2-(p_k+p_i-p_a)^2}.
\eq
As subtraction terms we use
\bq
\l s | V^{q_ag_i}_{k} | s' \r & = & 8 \pi \mu^{2\eps} \alpha_s C_F \delta_{ss'}
   \left[ \frac{2}{1-x+u} -(1+x) - \rho \eps (1-x) \right], \nonumber \\
\l s | V^{g_a \bar{q}_i}_{k} | s' \r & = & 8 \pi \mu^{2\eps} \alpha_s T_R \delta_{ss'}
     \left[ 
       1-\rho \eps - 2 x (1-x)
     \right], \nonumber \\
\l \mu | V^{q_aq_i}_{k} | \nu \r & = & 8 \pi \mu^{2\eps} \alpha_s C_F
     \left[ 
      -g^{\mu\nu} x + 4 \frac{(1-x)}{x} \frac{r(1-r)}{2p_iq} S^{\mu\nu}
     \right], \nonumber \\
\l \mu | V^{g_ag_i}_{k} | \nu \r & = & 16 \pi \mu^{2\eps} \alpha_s C_A
     \left[ 
      -g^{\mu\nu} \left( \frac{1}{1-x+u} -1 + x(1-x) \right) \right. \nonumber \\
      & & \left.
      + 2 (1-\rho \eps) \frac{(1-x)}{x} \frac{r(1-r)}{2p_iq} S^{\mu\nu}
     \right],
\eq
with
\bq
S^{\mu\nu} & = & 
\left( \frac{1}{r} p_i^\mu - \frac{1}{1-r} q^\mu \right)
\left( \frac{1}{r} p_i^\nu - \frac{1}{1-r} q^\nu \right).
\eq
Here $q$ is an arbitrary null vector not equal to $p_i$ and
\bq
r & = & \frac{2p_ip_a}{2p_ip_a+2qp_a}.
\eq
The momenta are mapped as follows:
\bq
\tilde{p}_a & = & x p_a, \nonumber \\
\tilde{p}_k & = & p_k + p_i - (1-x) p_a.
\eq
This mapping satisfies momentum conservation
\bq
\tilde{p}_k - \tilde{p}_a & = & p_k + p_i - p_a,
\eq
and the on-shell conditions
\bq
\tilde{p}_a^2 = 0, & & \tilde{p}_k^2 = m_k^2.
\eq
The desired asymptotic behaviour
\bq
\lim\limits_{u \rightarrow 0} \tilde{p}_a = x p_a,
& &
\lim\limits_{u \rightarrow 0} \tilde{p}_k = p_k
\eq
is fulfilled. 
In addition we have
\bq
S_{\mu\nu} \tilde{p}_a^\nu & = & 0,
\eq
e.g. $\tilde{p}_a$ is orthogonal to the spin correlation tensor.
We start with the integration of $V^{q_ag_i}_{k}$.
We obtain
\bq
{\cal V}^{q_ag_i}_{k}  & = & \int d\phi_{dipole} \frac{1}{2p_ip_a} V^{q_ag_i}_{k} 
 = \frac{\alpha_s}{2\pi} \frac{1}{\Gamma(1-\eps)} \left( \frac{4 \pi \mu^2 }{m_k^2-P^2} \right)^\eps {\cal V}^{qq}(x,x_0,\eps), \nonumber \\
\lefteqn{ {\cal V}^{qq}(x,x_0,\eps) } & & \nonumber \\
& = & x^\eps (1-x)^{-2\eps} (1-x_0 x)^{\eps} \int\limits_0^1 dw w^{-\eps-1} (1-w)^{-\eps} 
 \nonumber \\
 & & C_F \delta_{ss'} \left[ \frac{2}{1-x+u} -(1+x) -\rho \eps(1-x) \right] \nonumber \\
& = & C_F \delta_{ss'} x^\eps (1-x)^{-2\eps} (1-x_0 x)^{\eps}
 \frac{\Gamma(-\eps) \Gamma(1-\eps)}{\Gamma(1-2\eps)} \nonumber \\
& & \cdot \left(
     \frac{2}{1-x} {}_2F_1\left( 1, -\eps; 1-2\eps; \frac{-1}{1-x_0 x} \right)
     - (1+x) - \rho \eps (1-x) \right).
\eq
${\cal V}^{qq}$ is a distribution in $x$. In order to obtain expressions which are integrable
at $x=1$ we rewrite
\bq
(1-x)^{-2\eps-1} & = & \left. (1-x)^{-2\eps-1} \right|_+
+ \delta(1-x) \int\limits_0^1 dy (1-y)^{-2\eps-1},
\eq
where
\bq
\left. (1-x)^{-2\eps-1} \right|_+ 
& = & \left. \frac{1}{1-x} \right|_+ -2\eps \left. \frac{\ln(1-x)}{1-x} \right|_+ + O(\eps^2).
\eq
In order to expand the hypergeometric function we use the analytical continuation
formula \cite{Erdelyi}
\bq
{}_2F_1(a,b;c;-x) & = & \left(x\right)^{-a} \frac{\Gamma(c)\Gamma(b-a)}{\Gamma(b)\Gamma(c-a)}
 {}_2F_1\left(a,1+a-c;1+a-b;-\frac{1}{x}\right) \nonumber \\
 & & + \left(x \right)^{-b} \frac{\Gamma(c) \Gamma(a-b)}{\Gamma(a) \Gamma(c-b)}
 {}_2F_1\left(b,1+b-c;1+b-a;-\frac{1}{x} \right), \nonumber \\
& & \mbox{\hspace{6.5cm}} \left| \mbox{arg} x \right| < \pi.
\eq
We obtain
\bq
\lefteqn{ {\cal V}^{qq}(x,x_0,\eps) } & & \nonumber \\
& = & C_F \delta_{ss'} \left[
 -\frac{1}{\eps} \left( \left. \frac{2}{1-x} \right|_+ - (1+x) \right) + \rho(1-x) 
 \right. \nonumber \\
 & & 
 + \delta(1-x) \left( \frac{1}{\eps^2} + \frac{1}{\eps} \ln(2-x_0) + \frac{\pi^2}{6}
                      + 2 \ln(1-x_0) \ln(2-x_0) + 2 \; \mbox{Li}_2(x_0-1) \right. \nonumber \\
 & & \left. 
                      - \frac{1}{2} \ln^2(2-x_0) \right) 
 -2 \left( \ln x + \ln(2-x_0x) \right) \left(\left. \frac{1}{1-x} \right|_+ \right) 
 + 4 \left. \frac{\ln(1-x)}{1-x} \right|_+ \nonumber \\
 & & 
 \left. +(1+x) \left( \ln x-2\ln(1-x) + \ln(1-x_0x) \right) \right]
 + O(\eps).
\eq
For the integration of the other dipole terms we proceed in a similar way and we obtain
\bq
{\cal V}^{g_a\bar{q}_i}_{k}  & = & \int d\phi_{dipole} \frac{1}{2p_ip_a} \frac{n_s(q)}{n_s(g)} V^{g_a\bar{q}_i}_{k} 
 = \frac{\alpha_s}{2\pi} \frac{1}{\Gamma(1-\eps)} \left( \frac{4 \pi \mu^2 }{m_k^2-P^2} \right)^\eps {\cal V}^{gq}(x,x_0,\eps), \nonumber \\
\lefteqn{ {\cal V}^{gq}(x,x_0,\eps) } & & \nonumber \\
& = & x^\eps (1-x)^{-2\eps} (1-x_0 x)^{\eps} \int\limits_0^1 dw w^{-\eps-1} (1-w)^{-\eps} 
 T_R \delta_{ss'} \frac{1-\rho \eps -2x(1-x) }{1-\rho \eps} \nonumber \\
& = & T_R \delta_{ss'} x^\eps (1-x)^{-2\eps} (1-x_0 x)^{\eps}
 \frac{\Gamma(-\eps) \Gamma(1-\eps)}{\Gamma(1-2\eps)} 
 \frac{1-\rho \eps -2x(1-x)}{1-\rho\eps} \nonumber \\
& = &
 T_R \delta_{ss'} \left[
   -\frac{1}{\eps} \left( x^2+(1-x)^2\right) + 2 \rho x(1-x) 
   - \left( x^2+(1-x)^2 \right) \left( \ln x - 2 \ln(1-x) \right. \right. \nonumber \\
 & & \left. \left. 
 + \ln(1-x_0x) \right) \right]
 \nonumber \\
 & & + O(\eps),
\eq
\bq
{\cal V}^{q_aq_i}_{k}  & = & \int d\phi_{dipole} \frac{1}{2p_ip_a} \frac{n_s(g)}{n_s(q)} V^{q_aq_i}_{k} 
 = \frac{\alpha_s}{2\pi} \frac{1}{\Gamma(1-\eps)} \left( \frac{4 \pi \mu^2 }{m_k^2-P^2} \right)^\eps {\cal V}^{qg}(x,x_0,\eps), \nonumber \\
\lefteqn{ {\cal V}^{qg}(x,x_0,\eps) } & & \nonumber \\
& = & x^\eps (1-x)^{-2\eps} (1-x_0 x)^{\eps} \int\limits_0^1 dw w^{-\eps-1} (1-w)^{-\eps} 
 \nonumber \\
 & & C_F \left[ -g^{\mu\nu} x + 4 \frac{1-x}{x} \frac{r(1-r)}{2p_iq} S^{\mu\nu} \right] (1-\rho\eps) \nonumber \\
& = & C_F \left( -g^{\mu\nu} \right) \frac{\Gamma(-\eps)\Gamma(1-\eps)}{\Gamma(1-2\eps)}
x^\eps (1-x)^{-2\eps} (1-x_0x)^\eps \left( x + \frac{2}{1-\rho\eps} \frac{1-x}{x} \right) 
 (1-\rho\eps) \nonumber \\
& & + \mbox{gauge terms} \nonumber \\
& = & C_F \left(-g^{\mu\nu} \right) \left[
 - \frac{1}{\eps} \left( x + 2 \frac{1-x}{x} \right) + \rho x
   -\left( x + 2 \frac{1-x}{x} \right) \left( \ln x - 2 \ln(1-x) \right. \right. \nonumber \\
 & & \left. \left. 
 + \ln(1-x_0x) \right) 
 \right] \nonumber \\
& & + \mbox{gauge terms} \; + O(\eps),
\eq
\bq
{\cal V}^{g_ag_i}_{k}  & = & \int d\phi_{dipole} \frac{1}{2p_ip_a} V^{g_ag_i}_{k} 
 = \frac{\alpha_s}{2\pi} \frac{1}{\Gamma(1-\eps)} \left( \frac{4 \pi \mu^2 }{m_k^2-P^2} \right)^\eps {\cal V}^{gg}(x,x_0,\eps), \nonumber \\
\lefteqn{ {\cal V}^{gg}(x,x_0,\eps) } & & \nonumber \\
& = & x^\eps (1-x)^{-2\eps} (1-x_0 x)^{\eps} \int\limits_0^1 dw w^{-\eps-1} (1-w)^{-\eps} 
 \nonumber \\
 & & C_A \left[ -g^{\mu\nu} \left( \frac{1}{1-x+u} -1+x-x^2 \right)
                + 2 (1-\rho\eps) \frac{1-x}{x} \frac{r(1-r)}{2p_iq} S^{\mu\nu} \right] \nonumber \\
 & = & 2 C_A \left( -g^{\mu\nu} \right) x^\eps (1-x)^{-2\eps} (1-x_0x)^\eps
     \frac{\Gamma(-\eps)\Gamma(1-\eps)}{\Gamma(1-2\eps)} \nonumber \\
 & & \cdot \left[ \frac{1}{1-x} {}_2F_1\left(1,-\eps;1-2\eps;\frac{-1}{1-x_0x}\right)
                  -1+x-x^2 + \frac{1-x}{x} \right] \nonumber \\
 & & 
     + \mbox{gauge terms} \nonumber \\
 & = & 2 C_A \left( -g^{\mu\nu} \right)
  \left[ -\frac{1}{\eps} \left. \frac{1}{1-x} \right|_+ 
         - \left( \ln x + \ln(2-x_0x) \right) \left( \left. \frac{1}{1-x} \right|_+ \right)
         + 2 \left. \frac{\ln(1-x)}{1-x} \right|_+ 
  \right. \nonumber \\
  & & \left.
         + \left( \frac{1}{\eps} + \ln x - 2 \ln(1-x) + \ln (1-x_0x) \right)
           \left( 1 -x+x^2-\frac{1-x}{x} \right)
  \right. \nonumber \\
  & & \left.
         +\delta(1-x) \left( \frac{1}{2\eps^2} + \frac{\pi^2}{12} + \frac{1}{2\eps} \ln(2-x_0)
                             +\ln(1-x_0)\ln(2-x_0) + \mbox{Li}_2(x_0-1) 
  \right. \right. \nonumber \\
  & & \left. \left. 
                             -\frac{1}{4} \ln^2(2-x_0)
                      \right)
  \right] 
 + \mbox{gauge terms} \; + O(\eps).
\eq
Here $n_s(q)=2$ denotes the polarizations of a fermion and $n_s(g)=2(1-\rho\eps)$ denotes
the polariztions of a gluon. 
Note that the dependece on the momentum $q$ in ${\cal V}^{q_aq_i}_{k}$ and 
${\cal V}^{g_ag_i}_{k}$
dropped out after integration.

\subsection{Massive final emitter, massless initial spectator}

We now consider the case of a massive emitter (particle $k$) in the final state
and a massless spectator (particle $a$) in the initial state.
It is sufficient to consider the case where a heavy quark (or antiquark) emits a gluon.
The variables $z$ and $x$ of Catani and Seymour are given by
\bq
z = \frac{2p_kp_a}{2p_ip_a+2p_kp_a}, & & x = \frac{2p_ip_a+2p_kp_a-2p_ip_k}{2p_ip_a+2p_kp_a}.
\eq
The variables $z$ and $w$ are related by
\bq
z & = & 1 - \frac{1-x}{1-x_0x} w, \;\;\; x_0 = 1- \frac{m_k^2}{m_k^2-(p_k+p_i-p_a)^2}.
\eq
Since there is no collinear singularity, we just have to match the part of the soft singularity which
corresponds to this dipole factor:
\bq
\l s | V_{q_kg_i}^a | s' \r & = & 8 \pi \mu^{2\eps} \alpha_s C_F \delta_{ss'}
\left[ \frac{2}{1-z +(1-x)}  -2 (1-x_0) \frac{x^2}{1-x} \right] 
\eq
We use the same mapping for the momenta as in the previous section:
\bq
\tilde{p}_a & = & x p_a, \nonumber \\
\tilde{p}_k & = & p_k + p_i - (1-x) p_a.
\eq
This mapping fulfills the asymptotic behaviour
\bq
\lim\limits_{2p_kp_i \rightarrow 0} x = 1, \;\;\;
\lim\limits_{2p_kp_i \rightarrow 0} \tilde{p}_k = p_k + p_i, \;\;\;
\lim\limits_{2p_kp_i \rightarrow 0} \tilde{p}_a = p_a.
\eq
Integration yields
\bq
{\cal V}_{q_kg_i}^{a}  & = & \int d\phi_{dipole} \frac{1}{2p_ip_k} V_{q_kg_i}^{a} 
 = \frac{\alpha_s}{2\pi} \frac{1}{\Gamma(1-\eps)} \left( \frac{4 \pi \mu^2 }{m_k^2-P^2} \right)^\eps {\cal V}_{Qg}(x,x_0,\eps), \nonumber \\
\lefteqn{ {\cal V}_{Qg}(x,x_0,\eps) } & & \nonumber \\
& = & x^\eps (1-x)^{-2\eps} (1-x_0 x)^{\eps-1} \int\limits_0^1 dw w^{-\eps} (1-w)^{-\eps} 
 \nonumber \\
 & &
 C_F \left[ \frac{2}{1-z+(1-x)} -2(1-x_0) \frac{x^2}{1-x} \right] \nonumber \\
& = & C_F \delta_{ss'} x^\eps (1-x)^{-2\eps} (1-x_0 x)^{\eps-1}
 \frac{\Gamma(1-\eps)^2}{\Gamma(2-2\eps)} \nonumber \\
& & \cdot \left(
     \frac{2}{1-x} {}_2F_1\left( 1, 1-\eps; 2-2\eps; \frac{-1}{1-x_0 x} \right)
     -2(1-x_0) \frac{x^2}{1-x} \right) \nonumber \\
& = & 
 C_F \delta_{ss'} \left[
   \delta(1-x) \left( \frac{1}{\eps} \left( 1 + \ln(1-x_0) - \ln(2-x_0) \right)
                      + 2 - \frac{\pi^2}{3} + \ln(1-x_0) \right. \right. \nonumber \\
 & & \left. \left.
                      + \frac{1}{2} \ln^2(1-x_0) + \frac{1}{2} \ln^2(2-x_0)
                      - 2 \ln(1-x_0) \ln(2-x_0) - 2 \; \mbox{Li}_2(x_0-1)
               \right) \right. \nonumber \\ 
&  & \left. + 2 \left( \ln(2-x_0x) -\ln(1-x_0x) - \frac{(1-x_0)x^2}{1-x_0 x} \right) 
           \left( \left. \frac{1}{1-x} \right|_+ \right)
    \right] \nonumber \\
 & & 
 + O(\eps).
\eq

\section{Conclusions}

In this paper we have extended the dipole formalism to processes involving heavy fermions.
We gave the explicit subtraction terms, together with a mapping of the momenta
from the $(n+1)$-parton configuration to the $n$-parton configuration.
We evaluated the integrals of the subtraction terms over the dipole phase space
to order $O(\eps^0)$.
These ingredients are sufficient to set up numerical NLO programs based on the dipole
formalism for processes
involving heavy fermions.
An application to single-top production is in preparation \cite{We}.
Furthermore we used a systematic approach for the integration and subsequent expansion
in $\eps$. This method allows us in principle to obtain higher orders in $\eps$.
Our results are naturally expressed in terms of harmonic polylogarithms.
We are therefore confident that the techniques developed in this paper will prove useful also
in the cancellation of infrared singularities at next-to-next-to-leading order (NNLO).

\subsection*{Acknowledgments}

We would like to thank E. Laenen for discussions and useful comments on the manuscript.


\begin{appendix}

\end{appendix}

\end{document}